\documentclass{ws-mpla}
\newcommand{\tg}{{\rm tg\,}}
\newcommand{\ctg}{{\rm ctg\,}}
\newcommand{\bea}{\begin{eqnarray}}
\newcommand{\eea}{\end{eqnarray}}
\newcommand{\xv}{\vec{\rm x}}
\newcommand{\yv}{\vec{\rm y}}
\newcommand{\kv}{\vec{\rm k}}
\newcommand{\Jv}{\vec{\rm J}}
\newcommand{\hv}{\vec{\rm h}}
\newcommand{\uv}{\vec{\rm u}}
\newcommand{\pv}{\vec{\rm p}}
\newcommand{\Sv}{\vec{\rm S}}
\newcommand{\Mv}{\vec{\rm M}}
\newcommand{\Ev}{\vec{\rm E}}
\newcommand{\piv}{\vec{\pi}}
\newcommand{\nv}{\vec{\rm n}}
\newcommand{\Av}{\vec{\rm A}}
\newcommand{\ev}{\vec{\rm e}}
\newcommand{\etav}{\vec{\eta}}
\newcommand{\Bv}{\vec{\rm B}}
\newcommand{\xvp}{\vec{\rm x}_{\bot}}
\newcommand{\eijk}{\epsilon_{ijk}} 
\newcommand{\Lc}{{\cal L}}
\newcommand{\h}{\hbar}
\newcommand{\Lv}{\vec{\rm L}}
\newcommand{\Lcv}{\vec{\cal L}}   
\newcommand{\Jcv}{{\vec{{\cal{J}}}}}
\newcommand{\Jvv}{\vec{J}}
\def\bldmth#1{%
\mathchoice
{{\hbox{\boldmath$\displaystyle#1$\unboldmath}}}%
{{\hbox{\boldmath$\textstyle#1$\unboldmath}}}%
{{\hbox{\boldmath$\scriptstyle#1$\unboldmath}}}%
{{\hbox{\boldmath$\scriptscriptstyle#1$\unboldmath}}}%
}
\def\vec#1{\bldmth{#1}}
\begin{document}

\markboth{S.E.~Korenblit \& Kieun Lee}
{Disappearance of Schwinger's String at the Charge - Monopole ``Molecule''}

%%%%%%%%%%%%%%%%%%%%% Publisher's Area please ignore %%%%%%%%%%%%%%
\catchline{}{}{}{}{}
%%%%%%%%%%%%%%%%%%%%%%%%%%%%%%%%%%%%%%%%%%%%%%%%%%%%%%%%%%%%%%%%%%%

\title{DISAPPEARANCE OF SCHWINGER'S STRING\\AT THE CHARGE - MONOPOLE ``MOLECULE''}

\author{\footnotesize S.E.~KORENBLIT and KIEUN~LEE}

\address{
Department of Physics, Irkutsk State University,\\
Gagarin blvd 20, Irkutsk 664003, Russia\\
korenb@ic.isu.ru
}

\maketitle

\pub{Received (Day Month Year)}{Revised (Day Month Year)}

\begin{abstract}
An equivalence of total angular momentum operator of charge - monopole system to the 
momentum operator of a symmetrical quantum top is observed. This explicitly shows the 
string independence of Dirac's quantization condition leading to disappearance of 
Schwinger's string and reveals some properties of diatomic molecule for this system.
\end{abstract}

\ccode{PACS Nos.: 14.80.Hv}

\section{Introduction}	

It is well known\cite{Gold,Boul} that the one-particle wave function of a charge 
scattered on monopole's field decomposes as a sum of rotation eigenfunctions of 
symmetrical quantum top. Here some physical reasons of this will be elucidated. 

Let us remind the main results about the problem under consideration 
(see Ref. 3, Ref. 4, %%% \cite{Milt}, \cite{Cole} 
and references therein). It is well known that magnetic 
field's ``hedgehog'' $\Bv_m(\xv)$ of an infinitely heavy magnetic monopole placed at the 
origin $O$ and corresponded to residual (Abelian) $U(1)$ gauge group has a source at the 
origin $O$ viewed from classical point either as the origin of a semi-infinite 
infinitely thin solenoid (bar magnet string) $\Av^\pm_{\uv}(\xv)$ along the direction of 
vector ${\vec{\rm u}}$ (Dirac)\cite{Dirac}, or as the origin of two such symmetric 
strings $\Av_{\uv}(\xv)$ (Schwinger)\cite{Schw}: 
\bea
&&\!\!\!\!\!\!\!\!\!\!\!\!\!\!\!\!\!\!
\Av^\pm_{\uv}(\xv)=\frac{g}{r}\,\frac{(\uv\times\xv)}{\left((\uv\cdot\xv)\pm r\right)},
\quad \Av_{\uv}(\xv)=\frac 12\left(\Av^+_{\uv}(\xv)+\Av^-_{\uv}(\xv)\right),
\label{1} \\
&&\!\!\!\!\!\!\!\!\!\!\!\!\!\!\!\!\!\!
\Bv_m(\xv)=\left(\vec{\nabla}_{\rm x}\times\Av_{\uv}(\xv)\right)-\hv_{\uv}(\xv)
=g\frac{\xv}{r^3}, \quad r=|\xv|,
\label{2} \\
&&\!\!\!\!\!\!\!\!\!\!\!\!\!\!\!\!\!\!
\hv_{\uv}(\xv)=-2\pi{g}\uv\frac{(\uv\cdot\xv)}{r}\delta_{2\uv}(\xvp), 
\quad \xvp=\xv-\uv(\uv\cdot\xv),
\label{3} \\
&&\!\!\!\!\!\!\!\!\!\!\!\!\!\!\!\!\!\!
\mbox {for which: }\;\left(\vec{\nabla}_{\rm x}\cdot \Av^\pm_{\uv}(\xv)\right)=
\left(\vec{\nabla}_{\rm x}\cdot \Av_{\uv}(\xv)\right)=0,
\label{3_1} \\
&&\!\!\!\!\!\!\!\!\!\!\!\!\!\!\!\!\!\!
\left(\vec{\nabla}_{\rm x}\cdot\Bv_{m}(\xv)\right)=
-\left(\vec{\nabla}_{\rm x}\cdot\hv(\xv)\right)=4\pi g\delta_3(\xv),
\label{3_0}
\eea
where $\hv_{\uv}(\xv)$ - is the magnetic field inside the string, defined by 
two-dimensional $\delta_{2\uv}(\xvp)$ -- function. The same magnetic field (\ref{2}) 
results\cite{Milt} also from $\Av^\pm_{\uv}(\xv)$ with corresponding $\hv^\pm_{\uv}(\xv)$. 

In Cartesian basis $\ev_i$ placed at the origin $O$ for a charge position vector 
$\xv=r\nv=\rho{\etav}_{(\rho)}+z\ev_3$ one has the rotating vectors of spherical and 
polar bases ${\etav}_{(j)}(\beta,\alpha)$ as functions of corresponding angles 
$\alpha,\,\beta$:        
\bea
&&\!\!\!\!\!\!\!\!\!\!\!\!\!\!\!\!\!\!
\nv={\etav}_{(r)}=\ev_1\sin\beta\cos\alpha+\ev_2\sin\beta\sin\alpha+\ev_3\cos\beta,
\label{4} \\
&&\!\!\!\!\!\!\!\!\!\!\!\!\!\!\!\!\!\!
{\etav}_{(\beta)}=\ev_1\cos\beta\cos\alpha+\ev_2\cos\beta\sin\alpha-\ev_3\sin\beta,
\label{5_0} \\
&&\!\!\!\!\!\!\!\!\!\!\!\!\!\!\!\!\!\!
{\etav}_{(\alpha)}=-\ev_1\sin\alpha+\ev_2\cos\alpha,
\label{5} \\
&&\!\!\!\!\!\!\!\!\!\!\!\!\!\!\!\!\!\!
{\etav}_{(\rho)}=\ev_1\cos\alpha+\ev_2\sin\alpha= 
{\etav}_{(\beta)}\cos\beta+\nv\sin\beta.
\label{5_1} 
\eea
Thus, the gauge $\uv=\ev_3={\ev_z}$, with $(\uv\cdot\nv)=\cos\beta$, 
$(\uv\times\nv)={\etav}_{(\alpha)}\sin\beta$, recasts the different fields of 
Eq. (\ref{1}) to the following: 
\bea
&&\!\!\!\!\!\!\!\!\!\!\!\!\!\!\!\!\!\!
\Av^+_{\uv}(\xv)={\etav}_{(\alpha)}\,\frac{g}{r}\, \tg\frac{\beta}2,
\quad \Av^-_{\uv}(\xv)= -{\etav}_{(\alpha)}\,\frac{g}{r}\, \ctg\frac{\beta}2, 
\label{1_260} \\
&&\!\!\!\!\!\!\!\!\!\!\!\!\!\!\!\!\!\!
\Av_{\uv}(\xv)= -{\etav}_{(\alpha)} \,\frac{g}{r}\,\ctg\beta. 
\label{1_280}
\eea
Here the first and second ones are for the semi-infinite (Dirac) strings along $-\ev_z$ 
and $\ev_z$ respectively, whereas the third one is for the infinite (Schwinger) string, 
composed symmetrically by the two previous ones\cite{Milt}.

Various expressions (\ref{1}) for $\Av_{\uv}(\xv)$ are differed by the gauge 
transformation 
containing a multivalued gauge function\cite{Cole} $\Lambda(\xv)$. For example for  
transfer (rotation) from the semi-infinite string along $-\ev_z$ of Eq. (\ref{1_260}) 
to the one along the $\ev_z$ this transformation\cite{Milt,Cole} is a 
gauge one only ``almost everywhere'', out of the semi-infinite half plane, $y=0$, $x>0$, 
bounded by infinite $z$ axis. The potentials (\ref{1_260}) lead to\cite{Milt}:
\bea
&&\!\!\!\!\!\!\!\!\!\!\!\!\!\!\!\!\!
\frac{e}{c\h}\left(\Av^+_{\uv}(\xv)-\Av^-_{\uv}(\xv)\right)
=\vec{\nabla}_{\rm x}\Lambda(\xv)=
\frac{2Q}{\h}\frac{{\etav}_{(\alpha)}}{r\sin\beta},\; \mbox{ for:}
\label{0_290} \\
&&\!\!\!\!\!\!\!\!\!\!\!\!\!\!\!\!\!
\vec{\nabla}_{\rm x}=\nv\frac{\partial}{\partial r}+
\frac{{\etav}_{(\beta)}}{r}\frac{\partial}{\partial\beta}
+\frac{{\etav}_{(\alpha)}}{r\sin\beta} \frac{\partial}{\partial\alpha},
\quad \Lambda(\xv)=\frac{2Q}{\h}\alpha, 
\label{0_29} \\
&&\!\!\!\!\!\!\!\!\!\!\!\!\!\!\!\!\!
\mbox {with: }\;Q=\frac{eg}c,\; \mbox { and give: }\;
\piv_+=e^{i\Lambda(\xv)}\piv_- e^{-i\Lambda(\xv)},
\label{0_291} \\
&&\!\!\!\!\!\!\!\!\!\!\!\!\!\!\!\!\!
\mbox {for: }\;\piv_\pm=\pv-\frac{e}{c}\Av^\pm_{\uv}(\xv),\quad 
\pv=-i\h\vec{\nabla}_{\rm x}.
\label{0_293}
\eea
Single-valuedness of $e^{i\Lambda(\xv)}$ for $\alpha\rightarrow\alpha+2\pi$  
imposes Dirac's quantization conditions\cite{Dirac}: 
\begin{equation}
e^{i\Lambda(\xv)}=e^{iN\alpha},\quad -2Q=\h\,N, \quad N=0,\pm 1,\pm 2,\ldots.  
\label{0_D}
\end{equation}
Schwinger's symmetrical string of Eq. (\ref{1_280}) seems believable to lead 
to a more restrictive condition\cite{Schw} with $N\mapsto 2N$ only. 
However this string possess another interpretation without such a restriction, as it 
will be shown below.  

On the other hand, a common classical electromagnetic field (EMF) composing by the 
magnetic field $\Bv_m(\yv)$ from the monopole at the origin $O$ and by the electric 
field $\Ev_e(\yv)$ from the scattered point charge at the position $\xv$, brings into 
their system an additional irremovable angular momentum $\Mv=-Q\nv$, 
associated with Poynting momentum density vector\cite{Gold,Milt}. In spite of the 
impossibility to assign this angular momentum to any one of these particles, 
it has inspired Goldhaber\cite{Gold} and others \cite{Boul,Milt} to interpret 
the value $-Q$ quantized by Eq. (\ref{0_D}) as a conserving projection onto the vector 
$\nv$ of some additional quantum spin $\Sv$ satisfying the relations: 
\bea
&&\!\!\!\!\!\!\!\!\!\!\!\!\!\!\!\!\!\!\!\!
(\Mv\cdot\nv)=-Q\longleftrightarrow (\Sv\cdot\nv), \quad (\Sv\cdot\nv)\longmapsto \h\mu,
\quad 2\mu=N. 
\label{0_211} \\
&&\!\!\!\!\!\!\!\!\!\!\!\!\!\!\!\!\!\!\!\!
[S_i,S_j]=i\h\eijk{S_k},\quad [\Sv,x_j]=[\Sv,p_j]=0,
\label{1_2_03} \\
&&\!\!\!\!\!\!\!\!\!\!\!\!\!\!\!\!\!\!\!\!
[(\Sv\cdot\nv),\piv_s]=0,\; \mbox{ where for }\; \piv_s:\;
\frac{e}{c}\Av_s(\xv)=-\,\frac{(\Sv\times\xv)}{r^2},
\label{1_20_3} \\
&&\!\!\!\!\!\!\!\!\!\!\!\!\!\!\!\!\!\!\!\!
\mbox{with: }\;
\frac ec \Bv_m(\xv)=\frac 1{i\h}\left(\piv\times\piv\right)
=\frac{e}{c}\left(\vec{\nabla}_{\rm x}\times\Av_s(\xv)\right)-
\label{1_201} \\
&&\!\!\!\!\!\!\!\!\!\!\!\!\!\!\!\!\!\!\!\!
-\frac{i}{\h}\left(\frac ec\Av_s(\xv)\times\frac ec\Av_s(\xv)\right)=
-(\Sv\cdot\nv)\frac{\xv}{r^3},
\label{0_200} 
\eea
-- instead of the string expression (\ref{2}). Thus, he has avoided the use of any 
strings of Eq. (\ref{1}). The spin quantization condition (\ref{0_211}) is equivalent to 
(\ref{0_D}) making these strings invisible. However, instead of the string 
potentials of Eq. (\ref{1}), he has arrived to the non-Abelian 
spin-potential\cite{Milt,Cole} 
$\Av_s(\xv)$ given by Eq. (\ref{1_20_3}), which also obeys Eq. (\ref{3_1}) but is 
connected with the string ones $\Av^\pm_{\uv}(\xv)$ by a spin rotation that is meaningful 
only on the eigenstates of the third spin-component\cite{Boul,Milt} $S_3$. 
A single spin is enoug\cite{Gold} to obtain Dirac's strings 
$\Av^\pm_{\uv}(\xv)$ only. To reproduce Schwinger's string $\Av_{\uv}(\xv)$ of 
Eq. (\ref{1_280}) it is 
necessary to take the operator $\Sv$ as a sum of two mutually commutative spin 
operators\cite{Milt}, 
$\Sv=\Sv_a+\Sv_b$. The final result is reached by using the unitary 
transformation\cite{Milt}: ${\rm U}={\cal U}^{-1}_{a}(\alpha,\beta,-\alpha)
{\cal U}^{-1}_{b}(\alpha,\beta-\pi,-\alpha)$ (clf. Eq. (\ref{2_452}) 
below), rotated the projections of these two spins on the vector $\nv$ into their             
projections on the vectors $\pm\ev_z$ respectively: 
$(\Sv_{a,b}\cdot\nv)\mapsto\pm (\Sv_{a,b}\cdot\ev_3)=\pm(\Sv_{a,b})_3$, and furnished by 
eigenvalue condition 
$(\Sv_a-\Sv_b)'_3=-Q$. A crown of this cumbersome construction\cite{Boul,Milt} 
is an impression that for $\Lcv_s\equiv(\xv\times\piv_s)$ and with the first substitution 
of Eq. (\ref{0_211}) the total angular momentum operator $\Jvv_s$ takes a simple -- 
``one-particle'' form\cite{Gold} with the usual orbital momentum $\Lv=(\xv\times\pv)$ 
and the spin $\Sv$: 
\begin{equation}
\Jvv_s=\Lcv_s-Q\,\nv\longleftrightarrow \Lcv_s+(\Sv\cdot\nv)\,\nv=
{\Lv}+\Sv \equiv{\Jcv},  
\label{1_202} 
\end{equation}
and that the above rotation converts it to the standard one\cite{Milt}: 
\bea
&&\!\!\!\!\!\!\!\!\!\!\!\!\!\!\!\!\!\!\!\!\!
{\rm U}{\piv_s}{\rm U}^{-1}=\piv=\pv+\frac{{\etav}_{(\alpha)}}{r}
\sin\beta\left(\frac{S'_{a3}}{1+\cos\beta}+\frac{S'_{b3}}{1-\cos\beta}\right),
\label{0_805} \\
&&\!\!\!\!\!\!\!\!\!\!\!\!\!\!\!\!\!\!\!\!\!
{\rm U}\Jcv {\rm U}^{-1}=\Jvv=\left(\xv\times\piv\right)+\nv(\Sv_a-\Sv_b)'_3.
\label{805} 
\eea
Here $S'_{a3}=0$ for the one of Dirac's string whereas for Schwinger's one 
$(\Sv_a+\Sv_b)'_3=0$. Since $(\Lv\cdot\nv)=(\Lcv\cdot\nv)=0$, one has  
$(\Sv\cdot\nv)=(\Jcv\cdot\nv)$, what also helps to convert the Hamiltonian operator 
$2mH_s=\piv_s^2$ into the usual form\cite{Gold}:
\begin{equation}
{\rm U}{\piv_s^2}{\rm U}^{-1}=\piv^2=2mH={\widetilde p}^2_r+\frac{\Lcv^2}{r^2}. 
\label{264} 
\end{equation}
Note that one of the summands in Eq. (\ref{1}) always disappears as 
$\nv\rightarrow\pm\uv$ for the charge leaving ``visible'' only singular contribution 
of the another one, which is equal to 1/2 of Dirac's string as a ``half'' of Schwinger's 
string. This qualitatively makes very smooth a physical difference between two types 
of these strings, because by using a gauge-like transformation (\ref{0_290}) an arbitrary 
position of (Schwinger's) string always can be directed almost along the vector $\nv$ of 
incident charge without violating ``Dirac's veto''\cite{Milt}. 

The aim of the present letter is to show that the above one-particle interpretation of the 
total angular momentum operator can be replaced naturally by its interpretation for 
some extended object leading to disappearance of Schwinger's string. 

\section{The Algebra of operators in the presence of monopole}

Let us consider the algebra of operators for the motion of a charge in the monopole 
fields (\ref{1}) with the Hamiltonian (\ref{264}) and local commutation 
relations\cite{Gold,Boul,Milt,Cole}:
\bea
&&\!\!\!\!\!\!\!\!\!\!\!\!\!\!\!\!\!\!\!\!\!
[x_i,x_j]=0,\quad [x_i,\pi_j]=i\hbar\delta_{ij}, \quad 
[\pi_i\,,\pi_j]=i\frac{e\hbar}{c}\epsilon_{ijk}(\Bv_m)_k,
\label{3_2} \\
&&\!\!\!\!\!\!\!\!\!\!\!\!\!\!\!\!\!\!\!\!\!
\frac 12 \epsilon_{ijk}\left[\pi_i\,,[\pi_j\,,\pi_k]\right]=
\frac{e\hbar^2}{c}\left({\vec{\nabla}}_{\rm x}\cdot\Bv_m(\xv)\right)
= 4\pi\hbar^2 Q\delta_3(\xv), 
\label{3_3} 
\eea
In terms of the operator $\Lcv=(\xv\times\piv)$ the equation of motion takes its 
classical form\cite{Gold}:
\bea
&&\!\!\!\!\!\!\!\!\!\!\!\!\!\!\!\!\!\!\!\!\!\!
2mi\hbar\dot{\piv}=[\piv,\piv^2]=
i\frac{e\hbar}{c}\biggl((\piv\times\Bv_m)-(\Bv_m\times\piv)\biggr)
=\frac 2i \h \frac Q{r^3}{\Lcv},  
\label{3_600} \\
&&\!\!\!\!\!\!\!\!\!\!\!\!\!\!\!\!\!\!\!\!\!\!
2mi\h\dot{{{\Lcv}}}=\left[{{\Lcv}},\piv^2\right]=Q\,2mi\h\dot{\nv}=Q\,[\nv,\piv^2].
\label{3_70} 
\eea
The substitution $-Q\longleftrightarrow (\Jvv\cdot\nv)$ instead of the first one of 
Eq. (\ref{0_211}) for the angular momentum operator $\Jvv={\Lcv}-Q\nv\longleftrightarrow 
{\Lcv}+(\Jvv\cdot\nv)\nv$ instead of the $\Jcv$ in Eq. (\ref{1_202}) 
gives\cite{Gold,Boul,Milt,Cole,Lip}: 
\bea
&&\!\!\!\!\!\!\!\!\!\!\!\!\!\!\!\!\!\!\!\!\!\!
2mi\h\dot{\Jvv}=[\Jvv,\piv^2]=0, \;\; [(\Jvv\cdot\nv),\piv]=0,\;\;
[(\Jvv\cdot\nv),\nv]=0,  
\label{3_10} \\
&&\!\!\!\!\!\!\!\!\!\!\!\!\!\!\!\!\!\!\!\!\!\!
[J_i,x_j]=i\h\eijk{x_k},\;\; [J_i,\pi_j]=i\h\eijk{\pi_k},\;\; [J_i,J_j]=i\h\eijk{J_k}.
\label{3_11}
\eea
Hence $\Jvv$ is a conserving total angular momentum operator for the extended system: 
``charge + monopole + common EMF'', though the last two formulas in Eq. (\ref{3_11}) are 
valid, strictly speaking, only outside of the string $\hv_{\uv}(\xv)$ producing 
an additional contribution\cite{Milt} from Eq. (\ref{3}). 
The first relation of Eq. (\ref{3_2}) together with the first and the last relation of 
Eq. (\ref{3_11}) form the algebra of Euclidean group E3 having $(\Jvv\cdot\nv)$ as 
Casimir operator\cite{Lip}, what leads to quantization condition (\ref{0_D}) without 
any reference to explicit form of the potential $\Av^\pm_{\uv}(\xv)$. 
At last Jackiw\cite{Jac} has showed recently that Jacobi commutator (\ref{3_3}) provides 
the condition (\ref{0_D}) under direct construction of extended object such as 
tetrahedron. 

Hurst\cite{Hurst} was probably the first who used the differential form instead of the 
spin one (\ref{0_211}) for the projection of the total angular momentum operator 
$(\Jvv\cdot\nv)$ in the case of Dirac's potentials (\ref{1_260}). 
Remaining however in the framework of the single-particle interpretation of operator 
$\Jvv^2$ containing an additional extension parameter $\mu$  (clf. after Eq. (\ref{_450}) 
below) and imposing the (essential) self-adjointness condition for this operator with 
Dirac's or Schwinger's string, he 
obtained the charge quantization rules (\ref{0_D}) from the boundary conditions for its 
eigenfunctions that in fact make these strings invisible. 

Let us examine the operator $\Jvv$ in more detail. Though, its initial expression is not 
a sum of angular momentum operators:  
\begin{equation}
\Jvv={\Lcv}-Q\nv=\left(\xv\times\left(\pv-\frac{e}{c}\Av_{\uv}(\xv)\right)\right)-Q\nv, 
\label{16_0} 
\end{equation}
the use of the basis vectors of spherical and polar coordinate systems defined in Eqs. 
(\ref{4})--(\ref{5_1}) and Schwinger's type of the vector potential of Eq. 
(\ref{1_280}) recasts $\Jvv$ into the following: 
\begin{equation}
\Jvv=i\h\left[\frac{{\etav}_{(\beta)} }{\sin\beta}\frac{\partial}{\partial{\alpha}}
-{\etav}_{(\alpha)}\frac{\partial}{\partial{\beta}}\right]
-\frac{{\etav}_{(\rho)}}{\sin\beta}Q.
\label{16_1}
\end{equation}
Note that the basis vectors used here are not fully mutually orthogonal. 
The main observation of this work, surprisingly still not explicitly mentioned 
in the literature (see Refs. 3,4,13 %%% \cite{Milt}, \cite{Cole} 
and references therein) is that Cartesian 
components of this expression with Lipkin's\cite{Lip} and Hurst's\cite{Hurst} 
substitutions both together: 
\begin{equation}
-Q\longleftrightarrow(\Jvv\cdot\nv)\longleftrightarrow 
-i\h\frac{\partial}{\partial\gamma}, 
\label{180}
\end{equation}
exactly coincide, as it may be easily seen, with the standard expressions for Cartesian  
components of a total angular momentum operator of rotating rigid body -- a top, in terms 
of its Euler angles $\alpha,\;\beta,\;\gamma$ as dynamical variables\cite{Bied,LL}: 
\bea
&&\!\!\!\!\!\!\!\!\!\!\!\!\!\!\!\!\!\!\!\!\!
J_1(\alpha,\beta,\gamma)=
i\h\left[\ctg\beta\cos\alpha\frac{\partial}{\partial{\alpha}}+
\sin\alpha\frac{\partial}{\partial{\beta}}-
\frac{\cos\alpha}{\sin\beta}\frac{\partial}{\partial{\gamma}}\right],
\label{43} \\
&&\!\!\!\!\!\!\!\!\!\!\!\!\!\!\!\!\!\!\!\!\!
J_2(\alpha,\beta,\gamma)=
i\h\left[\ctg\beta\sin\alpha\frac{\partial}{\partial{\alpha}}-
\cos\alpha\frac{\partial}{\partial{\beta}}-
\frac{\sin\alpha}{\sin\beta}\frac{\partial}{\partial{\gamma}}\right],
\label{44} \\
&&\!\!\!\!\!\!\!\!\!\!\!\!\!\!\!\!\!\!\!\!\!
J_3(\alpha,\beta,\gamma)=-i\h\frac{\partial}{\partial{\alpha}}, 
\label{45} 
\eea
what is in exact correspondence with the meaning of the value of $(\Jvv\cdot\nv)$ as a 
projection of the total angular momentum operator onto the rotating axis $\nv$. 
Keeping in mind the Eq. (\ref{16_1}), may be the most explicit demonstration would be: 
\begin{equation}
\Jvv(\alpha,\beta,\gamma)=
i\h\left[\frac{{\etav}_{(\beta)} }{\sin\beta}\frac{\partial}{\partial{\alpha}}
-{\etav}_{(\alpha)}\frac{\partial}{\partial{\beta}}
-\frac{{\etav}_{(\rho)}}{\sin\beta}\frac{\partial}{\partial{\gamma}} \right].
\label{16_2}
\end{equation}
The expressions (\ref{43})--(\ref{16_2}) have nothing to do with singularities of 
Schwinger's string (\ref{1_280}), leading to the usual operator of total angular momentum 
square\cite{Bied}: $\Jvv^2=J^2_1+J^2_2+J^2_3$, or:  
\bea
&&
\Jvv^2(\alpha,\beta,\gamma)=
(i\h)^2\left[\frac{1}{\sin\beta}\frac{\partial}{\partial{\beta}}
\left(\sin\beta\frac{\partial}{\partial{\beta}}\right)+\right.
\label{_450} \\
&&
\left.+\frac{1}{\sin^2\beta}\left(\frac{\partial^2}{\partial{\alpha^2}}-
2\cos\beta\frac{\partial}{\partial{\alpha}}\frac{\partial}{\partial{\gamma}}+
\frac{\partial^2}{\partial{\gamma}^2}\right)\right].
\nonumber
\eea
This expression was previously used for a charge - monopole 
system\cite{Boul,Milt,Cole,Hurst,Schw_m} only with fixed eigenvalues of the operators 
$J_3\mapsto\h m$, and/or $(\Jvv\cdot\nv)\mapsto\h\mu$. For the last case Hurst\cite{Hurst} 
gives also the expressions for the Cartesian components of $J_k$ (\ref{16_0}) with 
Dirac's string (\ref{1_260}) that however have nothing to do with Eq. (\ref{16_1}) and 
the top angular momentum (\ref{43})--(\ref{16_2}). 

In the Cartesian basis of fixed coordinate system ${\ev}_{i}$ the basis of the rotating 
coordinate system $\vec{f}_{(j)}$ tightly associating with the rotating top for 
$i,j=1,2,3$, has the following form\cite{Bied}:
\begin{equation}
\vec{f}_{(j)}(\alpha\beta\gamma)=\sum_{i=1}^3 R_{ij}(\alpha\beta\gamma){\ev}_{i},
\quad ({\ev}_{i}\cdot{\ev}_{j})=(\vec{f}_{(i)}\cdot\vec{f}_{(j)})=\delta_{ij}.
\label{53_0} 
\end{equation}
Conversion to this rotating system is realized by rotation matrix 
$\widehat{R}(\alpha\beta\gamma)$ with matrices elements 
$R_{ij}=({\ev}_{i}\cdot\vec{f}_{(j)})$ given in Ref. 10 %%% \cite{Bied}
depending on the Euler angles: $0\leq\alpha<2\pi$, $0\leq\beta\leq\pi$, 
$0\leq\gamma<2\pi$, 
\bea
\widehat{R}(\alpha\beta\gamma)=\left(\begin{array}{cccc}
\cos\alpha\cos\beta\cos\gamma &\;\,& -\cos\alpha\cos\beta\sin\gamma &\;\,
\cos\alpha\sin\beta 
\\
-\sin\alpha\sin\gamma &\;\,& -\sin\alpha\cos\gamma &
\\
-------&\;\,& -------&\;\,---- 
\\
\sin\alpha\cos\beta\cos\gamma &\;\,& -\sin\alpha\cos\beta\sin\gamma &\;\,
\sin\alpha\sin\beta 
\\
+\cos\alpha\sin\gamma &\;\,& +\cos\alpha\cos\gamma &
\\
-------&\;\,& -------&\;\,---- 
\\
-\sin\beta\cos\gamma &\;\, & \sin\beta\sin\gamma &\;\, \cos\beta
\end{array}
\right).
\label{m_22}
\eea
Because for $l=1,2,3$, the vectors $\vec{f}_{(l)}$ are vector operators with respect to 
rotation generated by operator $\Jvv$ (\ref{16_2}), the components of this differential 
operator in the rotating coordinate system are scalars $J_{(l)}\equiv{\cal{P}}_{(l)}$, 
defined as\cite{Bied}: 
\bea
&&\!\!\!\!\!\!\!\!\!\!\!\!\!\!\!\!\!\!\!\!\!
{\cal{P}}_{(l)}=(\vec{f}_{(l)}\cdot\Jvv)=\sum_{i=1}^3 R_{il}{J}_{i},
\;\mbox{ where: }\;[J_i,\,R_{il}]=0,
\label{54} \\
&&\!\!\!\!\!\!\!\!\!\!\!\!\!\!\!\!\!\!\!\!\!
[J_i,\,R_{j\,l}]=[J_i,\,(\vec{f}_{(l)})_j]=
i\h\eijk\,(\vec{f}_{(l)})_k=i\h\,\eijk\,{R_{kl}},
\label{54_01} \\
&&\!\!\!\!\!\!\!\!\!\!\!\!\!\!\!\!\!\!\!\!\!
{\cal{P}}_{(1)}=\Lc_{(1)}=
i\h\left[-\ctg\beta\cos\gamma\frac{\partial}{\partial{\gamma}}
-\sin\gamma\frac{\partial}{\partial{\beta}}
+\frac{\cos\gamma}{\sin\beta}\frac{\partial}{\partial{\alpha}}\right]\!,
\label{55} \\
&&\!\!\!\!\!\!\!\!\!\!\!\!\!\!\!\!\!\!\!\!\!
{\cal{P}}_{(2)}=\Lc_{(2)}=i\h\left[
\ctg\beta\sin\gamma\frac{\partial}{\partial{\gamma}}
-\cos\gamma\frac{\partial}{\partial{\beta}}
-\frac{\sin\gamma}{\sin\beta}\frac{\partial}{\partial{\alpha}}\right],
\label{56} \\
&&\!\!\!\!\!\!\!\!\!\!\!\!\!\!\!\!\!\!\!\!\!
{\cal{P}}_{(3)}=(\Jvv\cdot\nv)=-i\h\frac{\partial}{\partial{\gamma}},
\quad \Lc_{(3)}=(\nv\cdot\vec{\Lc})=0,
\label{57}
\eea
-- since the vector $\vec{f}_{(3)}\equiv \nv(\beta,\alpha)$ for all $\gamma$. These 
components obey the relations\cite{Bied} (* - means complex conjugation):
\bea
&&
[{\cal P}_{(i)},{\cal P}_{(j)}]=-i\h\eijk{\cal P}_{(k)},  \quad 
[J_i,{\cal P}_{(j)}]=0,
\label{57_01} \\
&&
{\cal P}_{(k)}(\alpha,\beta,\gamma)=J^*_k(-\gamma,-\beta,-\alpha). 
\label{57_03}
\eea
The operator (\ref{_450}) of  total angular momentum square coincides for the 
both coordinate systems\cite{Bied}. Therefore, the angle-dependent part $\Lcv^2/r^2$ of 
the Hamiltonian (\ref{264}) in fact represents the Hamiltonian of symmetric 
top\cite{Bied,LL} with an infinite moment of inertia about the third principal axis of 
inertia, $\vec{f}_{(3)}=\nv$. The projection ${\cal{P}}_{(3)}$ onto this axis of the 
total angular 
momentum $\Jvv$ then remains to be a constant $-Q$ and with $\vec{\cal{P}}^2=\Jvv^2$:
\begin{equation}
\Lcv^2=\Jvv^2-Q^2 \longleftrightarrow
\vec{\cal P}^2-{\cal P}^2_{(3)}={\cal P}^2_{(1)}+{\cal P}^2_{(2)}, 
\label{57_04}
\end{equation}
which can be considered as the main result of this work.

\section{The wave function and the scattering amplitude}

The above observations reveal a deep similarity between the rotation wave functions of 
the usual spinless diatomic molecule with taking into account the total 
orbital momentum of its electronic shell\cite{LL}, and the rotation wave functions 
of the effective ``molecule'' composed by the electric charge and magnetic monopole 
(or by the two dyons) with taking into account the angular momentum of their common EMF.  
Thus for the states with the fixed total angular momentum for both these ``molecules'', 
$\Jvv^2\mapsto\h^2 j(j+1)$, the states of the usual molecule with conserving projection 
of its averaged electronic shell orbital momentum onto the rotating molecule's axis $\nv$: 
$(\overline{\vec{\rm L}}_e\cdot\nv)=(\Jvv\cdot\nv)\mapsto\h\lambda$, 
with the obvious condition\cite{LL} $j\geq |\lambda|$, are in direct correspondence 
with the states of the charge-monopole ``molecule'' with a certain conserved projection 
of the angular momentum of their common EMF\cite{Milt,Cole} onto the rotating 
``molecule's'' axis $\nv$: $(\Sv\cdot\nv)=(\Jvv\cdot\nv)\mapsto\h\mu= -Q$, what 
eventually gives the one and the same conditions for the one and the same eigenfunctions 
of symmetric quantum top for the ``molecules'' of  both types, with the 
replacement\cite{Boul,Milt,LL}: $\lambda\leftrightarrow\mu$. 
For matrix representation ${\rm J}_k$ of angular momentum operators, with   
$J_k\mapsto\hbar{\rm J}_k$, these well known Wigner's $D$- functions\cite{Bied} appear 
from: 
\bea
&&\!\!\!\!\!\!\!\!\!\!\!\!\!\!\!\!\!\!
\left[{\rm J}_i\,,{\rm J}_j\right]=i\epsilon_{ijk}{\rm J}_k,\quad 
{\cal U}(\alpha,\beta,\gamma)=
e^{-i\alpha{\rm J}_3}e^{-i\beta{\rm J}_2}e^{-i\gamma{\rm J}_3}, 
\label{2_453_00} \\
&&\!\!\!\!\!\!\!\!\!\!\!\!\!\!\!\!\!\!
\frac 1{i^j}\sqrt{\frac{8\pi^2}{2j+1}}\,\langle\alpha\beta\gamma|jm\rangle_\mu=
\langle j\mu|\,{\cal U}^{-1}(\alpha,\beta,\gamma)|jm\rangle= 
\label{2_453_01} \\
&&\!\!\!\!\!\!\!\!\!\!\!\!\!\!\!\!\!\!
=D^{(j)*}_{m,\mu}(\alpha,\beta,\gamma)=
e^{i\mu\gamma}\,d^j_{m,\mu}(\cos \beta)\,e^{im\alpha},
\label{2_453_0} 
\eea
as the common eigenfunctions of the operators $\Jvv^2$, $J_3$, and 
$\Bigl(\Jvv\cdot\nv(\beta,\alpha)\Bigr)={\cal{P}}_{(3)}$ with the eigenvalues 
$\h^2 j(j+1)$, $\h m$, and $\h \mu$ respectively, for which $-j\leq m, \mu\leq j$. When 
$\gamma= -\alpha$, these eigenfunctions are reduced to\cite{Bied,LL}: 
\bea
&&\!\!\!\!\!\!\!\!\!\!\!\!\!\!\!
{\cal U}^{-1}(\alpha,\beta,-\alpha)=
e^{-i\alpha{\rm J}_3}e^{i\beta{\rm J}_2}e^{i\alpha{\rm J}_3}=
\exp\{i\beta\,(\Jv\cdot{\etav}_{(\alpha)})\},
\label{2_452} \\
&&\!\!\!\!\!\!\!\!\!\!\!\!\!\!\!
\frac 1{i^j}\sqrt{\frac{4\pi}{2j+1}}\,\langle\nv(\beta,\alpha)|jm\rangle_\mu=
\langle j\mu|\,{\cal U}^{-1}(\alpha,\beta,-\alpha)|jm\rangle, 
\label{2_451} \\
&&\!\!\!\!\!\!\!\!\!\!\!\!\!\!\!
\mbox {giving: }\;
\langle\alpha\beta\gamma|jm\rangle_\mu=\frac{e^{i\mu(\gamma+\alpha)}}{\sqrt{2\pi}}\,
\langle\nv(\beta,\alpha)|jm\rangle_\mu.
\label{2_454} 
\eea
Thus, for the gauge $\uv={\ev_z}$ Schwinger's string (\ref{1_280}) is ``dissolved'' in 
the total angular momentum operator for ``charge + monopole + common EMF'' if the latter 
is considering as a total angular momentum operator of some effective extended quantum 
object with the properties of the symmetric top. So, the above eigenvalues $j,m,\mu$ 
of the mutually commutative (differential) operators $\Jvv^2$, $J_3$, and 
${\cal{P}}_{(3)}$, can be integer as well as half integer. 
Indeed, unlike usual diatomic molecules, the additional common EMF of charge and monopole 
should not induce here only purely integer orbital momentum, whereas the disappearance of 
the string makes irrelevant Schwinger's narrowing\cite{Schw} onto the even $N$. 
Note that the notion of extended (impenetrable rigid) body in quantum mechanics 
admits both integer and half integer values of its angular momentum\cite{Bied}.  

When the charge falls along the $z$ axis from $z=-\infty$, one has $\nv= -\ev_z$ and for 
the eigenvalue $\h m$ of the operator $J_3=(\Jvv\cdot\ev_z)\longmapsto-(\Jvv\cdot\nv)$ 
obtains\cite{Boul,Milt} $m=-\mu$. Thus, the exact scattering wave function 
$\psi^{(+)}_{\kv}(\xv)$ and the full scattering amplitude ${\cal F}(k^2,\cos\beta)$ are 
connected by the relation\cite{Gold,Milt}:
\bea
&&\!\!\!\!\!\!\!\!\!\!\!\!\!\!\!\!\!\!\!\!\!
\psi^{(+)}_{\kv}(\xv)=
e^{-i\pi\mu}\sum\limits^\infty_{j=|\mu|}(2j+1)e^{i\pi j}e^{-i\pi\ell/2}\,
{\rm j}_\ell(kr)\,D^{(j)*}_{-\mu,\mu}(\alpha,\beta,-\alpha), 
\label{6_S4} \\
&&\!\!\!\!\!\!\!\!\!\!\!\!\!\!\!\!\!\!\!\!\!
\psi^{(+)}_{\kv}(\xv)\underset{r\to\infty}{\longrightarrow} e^{-2i\mu\alpha}
\left[e^{i(\kv\cdot\xv)}+{\cal F}(k^2,\cos\beta)\frac{e^{ikr}}r\right],
\label{6_S7}  \\
&&\!\!\!\!\!\!\!\!\!\!\!\!\!\!\!\!\!\!\!\!\!
2ik {\cal F}(k^2,\cos\beta)=
e^{-i\pi\mu}\sum\limits^\infty_{j=|\mu|}(2j+1)e^{-i\pi(\ell-j)}
\,d^j_{-\mu,\mu}(\cos \beta),
\label{6_S8} 
\eea
what follows\cite{Milt} from asymptotic behavior of the Bessel function 
${\rm j}_\ell(kr)$:  
\begin{equation}
{\rm j}_\ell(kr)\underset{r\to\infty}{\longrightarrow}
\frac 1{kr}\sin\left(kr-\frac{\pi\ell}2\right),
\quad 
\ell+\frac 12=\left[\left(j+\frac 12\right)^2-\mu^2\right]^{1/2}\!\!\!.
\label{6_S6} 
\end{equation}
When in Eq. (\ref{6_S4}) the eigenfunctions of a symmetric top of Eq. (\ref{2_453_0}) is 
used, the multiplier $e^{-2i\mu\alpha}$ that deforms also the falling plane 
wave\cite{Boul,Milt} in expression (\ref{6_S7}), is replaced by 
$e^{i\mu(\gamma-\alpha)}$. However for the fixed $\mu$ the dependence on angle $\gamma$ 
coming here from Eq. (\ref{2_454}) gives only a common phase as well as for the case of 
the usual diatomic molecule\cite{LL}. Therefore, this multiplier has no physical meaning 
and can not change the one-particle interpretation of the scattering wave function 
(\ref{6_S4}) and scattering amplitude (\ref{6_S8}), because the vector $\nv$ in Eq. 
(\ref{4}) depends on Euler angles $\alpha,\,\beta$ only, where $\beta$ becomes a 
scattering angle. 

\section{Conclusion}  

The inability to assign irremovable additional angular momentum (\ref{0_211}) 
of common charge-monopole EMF to any one of these particles indicates incompleteness 
of single-particle interpretation\cite{Hurst} of the rotation symmetry for this 
system and lack of single-particle interpretation for the total angular momentum 
operator and its eigenfunctions.
We showed that Schwinger's symmetric vector potential\cite{Schw} (\ref{1}), (\ref{1_280}) 
directly leads to the more natural interpretation of the rotation symmetry 
for this system, as a symmetry of extended object with the properties of a symmetric 
quantum top with the infinite moment of inertia about the third principal axis of 
inertia. 
The whole system behaves with respect to rotations similarly to diatomic molecule 
with taking into account the total angular momentum of its electronic shell. 
Adjacent results with different interpretation via ``spinning-isospinning top''
were obtained in Ref. 12 %% \cite{Plyusch} 
starting from purely classical consideration of the charge-monopole system.

\section*{Acknowledgments}

The authors are grateful for useful discussions to A.N.~Vall, and also to S.V.~Lovtsov 
and A.E.~Rastegin.

This work was supported in part by the RFBR (project N 09-02-00749) and by the program 
``Development of Scientific Potential in Higher Schools'' (project N 2.2.1.1/1483, 
2.1.1/1539)

\end{document}